\documentclass[aip,apl,jmp,groupedaddress,amsmath,amssymb,reprint]{revtex4-1}
\usepackage[latin9]{inputenc}
\usepackage{textcomp}
\usepackage{amstext}
\usepackage{amssymb}
\usepackage{graphicx}

\makeatletter
 
 \@ifundefined{textcolor}{}
 {%
   \definecolor{BLACK}{gray}{0}
   \definecolor{WHITE}{gray}{1}
   \definecolor{RED}{rgb}{1,0,0}
   \definecolor{GREEN}{rgb}{0,1,0}
   \definecolor{BLUE}{rgb}{0,0,1}
   \definecolor{CYAN}{cmyk}{1,0,0,0}
   \definecolor{MAGENTA}{cmyk}{0,1,0,0}
   \definecolor{YELLOW}{cmyk}{0,0,1,0}
 }

\usepackage{dcolumn}
\usepackage{bm}

\makeatother

\begin{document}

\title{Aharonov-Bohm effect in an electron-hole graphene ring
system}

\author{D. Smirnov}
\email{smirnov@nano.uni-hannover.de}
\author{H. Schmidt}
\author{R. J. Haug}

\affiliation{Institut f\"ur Festk\"orperphysik, Leibniz Universit\"at Hannover, Appelstr. 2, 30167 Hannover, Germany}

\date{\today}
\begin{abstract}
Aharonov-Bohm oscillations are observed in a graphene quantum ring with a top gate 
covering one arm of the ring. As graphene is a gapless semiconductor this geometry allows to study not only the quantum interference of electrons with electrons or holes with holes but also  the unique situation of quantum interference between electrons and holes. The period and
amplitude of the observed Aharonov-Bohm oscillations are independent of the sign of the applied
gate voltage showing the equivalence between unipolar and dipolar interference.
\end{abstract}

\maketitle

One of the best known effects that can be used
to observe and control quantum interference is the 
Aharonov-Bohm (AB) effect \cite{AharonovBohmEffektFP1959[5],Exp4[9]}.
During the last years the AB effect was intensively studied for two-dimensional systems in semiconducting 
 heterostructures\cite{qdring1,qdring2,qdring3}. The introduction of graphene \cite{Erscheinungspaper2004[1]} 
opened new ways to study electronic and phase coherent transport in a two-dimensional system.
Therefore, several theoretical studies concerning the AB effect in graphene were 
published in recent years \cite{Theorie2[11],Theory1[10]}, 
but only very few experimental works were carried out \cite{Expr1[6],Exp2[7],Exp3[8]}.  One of the remarkable
effects in graphene is that both charge carrier types, electrons and holes, 
can be induced in one and the same sample with local gates \cite{BandTheoryOfGraphite[2], 
TG4[17],TGsusp[18]}.  In such experiments new effects, like the rise of the 
values of Quantum Hall plateaus  \cite{TG(QHETHEORIE),Kleintunnelingtheorie[20]}  
and the non-perfect Klein tunneling \cite{Falko2006,Katsnelson2006}  were observed. 
\newline
In this paper we present an experiment where it is possible to combine the AB-effect
and Klein tunneling. While using a local gate to create a pnp-junction
we are able to show the AB-effect not only for an unipolar system
but also for an electron-hole system. Such interference between electrons and holes can only be observed due to our special sample structure and due  to the unique bandstructure of graphene. 
\newline
The sample was fabricated via a standard procedure: the graphene flake
was produced by mechanical exfoliation from natural graphite and
deposited on a $285\,\mathrm{nm}$ thin layer of $\text{SiO}_{2}$
on top of a heavily p-doped silicon wafer, which was used as a backgate (BG)
during the measurements. The sample was found to be a monolayer graphene
flake by optical microscopy using the light intensity contrast
shift analysis method in the green channel \cite{blake:063124}.
Electron beam lithography and oxygen plasma etching were used to define
a ring with an inner radius of $220\,\mathrm{nm}$ and an outer radius of $360\,\mathrm{nm}$.
Figure 1 shows an  image
of the etched device. In a second step Chromium/Gold contacts were evaporated.
In the third step another layer of Polymethylmethacrylate (PMMA) was
deposited on top of the flake to enable the fabrication of a topgate (TG)
using the PMMA as an insulator \cite{TG1[14],schmidt:Paper,TG2[15]}.
The Chromium/Gold topgate was evaporated over one arm of the ring (see Fig.~1(b)).
Before the measurements, the sample was annealed for more than eight
hours with an average temperature of $250\text{\textdegree C}$ to
reduce doping and increase mobility.
\begin{figure}
\begin{centering}
\includegraphics{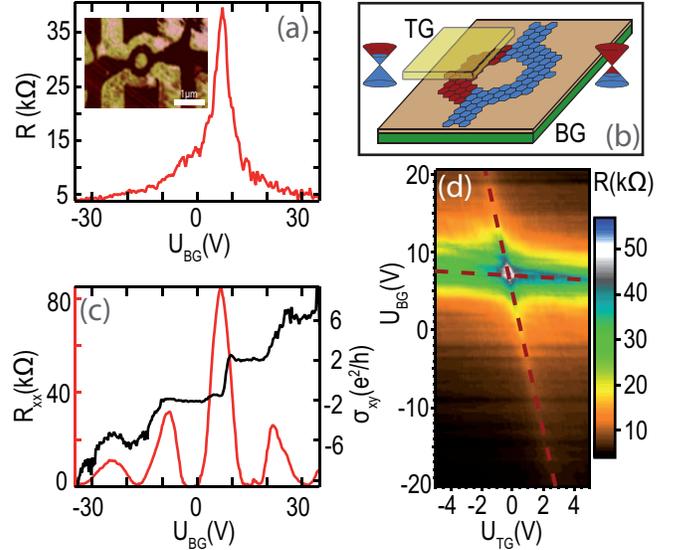} 
\par\end{centering}
\caption{\textit{\label{FIg1}(a) Four probe resistance measurements over the
ring versus backgate voltage. The inset shows an Atomic Force Microscope picture of the
sample. (b) Schematic picture of the graphene ring with different charge
carriers in the ring. (c) Longitudinal resistance and Hall conductivity
versus backgate voltage with a magnetic field of $13\,\mathrm{T}$ applied. (d) Resistance
measurements for different topgate and backgate voltages showing two
charge neutrality lines. }}
\end{figure}
 All measurements presented in
this paper are four probe measurements and have been performed in
a $\mathrm{He}^{3}$- cryostat with a base temperature of $500\,\mathrm{mK}$.
A perpendicular magnetic field of up to $13\,\text{T}$ was applied.
The resistance was measured with a lock-in amplifier with a current
of $5\,\mathrm{nA}$.

Figure~1(a) shows the measured resistance of the ring versus the backgate
voltage. The charge neutrality point (CNP) is observed at a gate voltage
of $7.25\,\text{V}$. We attribute this small but non-zero gate voltage
to doping that has not been removed through the annealing process
and to the extra layer of PMMA that was deposited on top of the sample.
The sample was identified as monolayer graphene also by magnetotransport
measurements which show the typical half integer Quantum Hall effect
\cite{HIQHE_THEO,QHE1[4]} (see in Fig.~1(c)). The mobility
for holes $\mu\sim6000\,\frac{\mathrm{cm^{2}}}{\mathrm{Vs}}$ and
for electrons $\mu\sim6800\,\frac{\mathrm{cm^{2}}}{\mathrm{Vs}}$
was calculated from the backgate dependent resistance measurements,
taking into account the geometric factor of the sample. The mean free
path is approx. $105\,\mathrm{nm}$ which is much smaller 
than the ring circumference $L=1.8\,\mathrm{\text{\textmu}m}$.
This means that the measured system is in the diffusive regime.

Such transport measurements are also used to characterize the topgate coupling.
Figure~1(d) shows  the colour intensity plot of the ring resistance as a function of backgate and topgate
voltage. One can clearly see both CNPs (indicated
by dashed red lines), which divide the color plot in four different
regions: two unipolar (electrons-electrons, holes-holes) and two bipolar
regions (electrons-holes, holes-electrons). The coupling factor $\alpha_{TG}=5.67\cdot\alpha_{BG}$
is in good agreement with the expected parallel capacity coupling
model based on the distance of $50\,\mathrm{nm}$ between flake and
topgate.

Figure~2 shows the AB effect measurements for a fixed backgate voltage
$U_{BG}=0\,\mathrm{V}$ and vanishing topgate voltage corresponding
to $p=5.7\cdot10^{15}\,\mathrm{m^{-2}}$. The magnetic field was swept
in a limited range around zero magnetic field in order to avoid the
occurrence of Shubnikov-de Haas oscillations. Figure~2(a) shows the ring
resistance as a function of magnetic field. A peak
can be observed at zero Tesla which is attributed to weak localization.
Small oscillations are seen over the whole shown magnetic-field range. These
oscillations have an average visibility of $0.3\,\%$ and can be identified
as AB oscillations. Figure~2(b) shows the AB oscillations with the
background subtracted. The background resistance was obtained by a
running average over a number of AB periods with a minimum of one
period \cite{Expr1[6],Exp2[7]}.

Figure~2(c) shows the Fourier spectrum of the oscillations presented in Fig.~2(b). The Fourier
spectrum has a peak
at $\Delta B^{-1}=62\,\mathrm{T^{-1}}$ which corresponds to a period
of $\Delta B_{AB}=16\,\mathrm{mT}$. The expected period for a ring $\Delta B_{AB}=h/({e}\pi{r}^2)$
with an average radius ${r}$ of $290\,\mathrm{nm}$ is $15.6\,\mathrm{mT}$,
so the measured oscillations match the first $h/e$ harmonic and fit
the size of the ring. The black curve is a Gaussian fitted 
to the Fourier spectrum. The curve illustrates the period of the oscillations
and the spreading which can be a hint towards the different paths
possible within the geometric width of the ring. In the Fourier spectrum
we observe a tail around $\Delta B^{-1}=125\,\mathrm{mT^{-1}}$,
which can be an indication towards the second harmonic and explains
the strong modulation of the oscillations. The phase coherence length of our system has to be shorter than
two times the ring circumference, since there appear only indications of a second harmonic.

\begin{figure}
\begin{centering}
\includegraphics{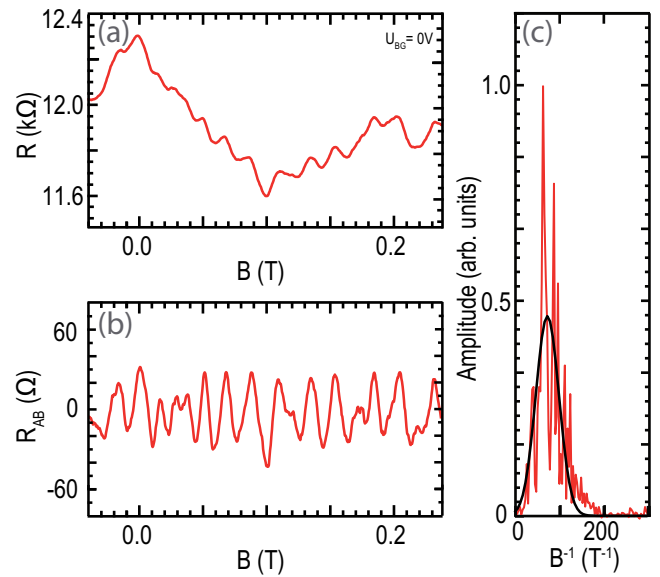} 
\par\end{centering}

\caption{\label{FIg2} \textit{Aharonov-Bohm oscillations: (a) Four probe resistance
measurements over the ring versus a perpendicular magnetic field at
a constant backgate voltage $U_{BG}=0\,\mathrm{V}$. (b) Same oscillations
with subtracted background resistance. The period of the oscillations
is $16,5\,\mathrm{mT}$. (c) Fourier spectrum of the oscillations (red)
and a Gaussian fit (black). }}
\end{figure}

AB measurements are reported with comparable results in Refs. \cite{Exp2[7],Exp3[8],Expr1[6]}. 
For comparison in Ref. \cite{Expr1[6]} oscillations were measured with a visibility of less than 1\% in
low magnetic fields which was attributed to a possible defect in one
arm of the ring. Other experiments showed AB oscillations
with a visibility of up to 5\% for a sample with a side gate \cite{Exp2[7]}.
In both experiments no second harmonics oscillations were observed
in low magnetic fields. Our observed visibility is comparable with
the results presented in \cite{Expr1[6]} but we cannot connect the
results with a defect in our sample.

In contrast to the previous experiments, our experimental set-up allows
the generation of different charge carriers in the two arms of the
ring by varying the topgate voltage. The topgate-dependent measurements
are performed as follows: Firstly the backgate voltage was fixed at
a certain value to define the carrier type and the concentration in
the leads and one arm of the ring. Secondly the topgate voltage was
set to define the charge carriers in the second arm of the ring. Thirdly
the magnetic field was swept and the voltage was measured and the
resistance calculated. The AB oscillations were obtained by subtracting
the background resistance as described before and the absolute amplitude of the resulting oscillations
are analyzed by the root mean square (RMS) value.
\begin{figure}
\begin{centering}
\includegraphics{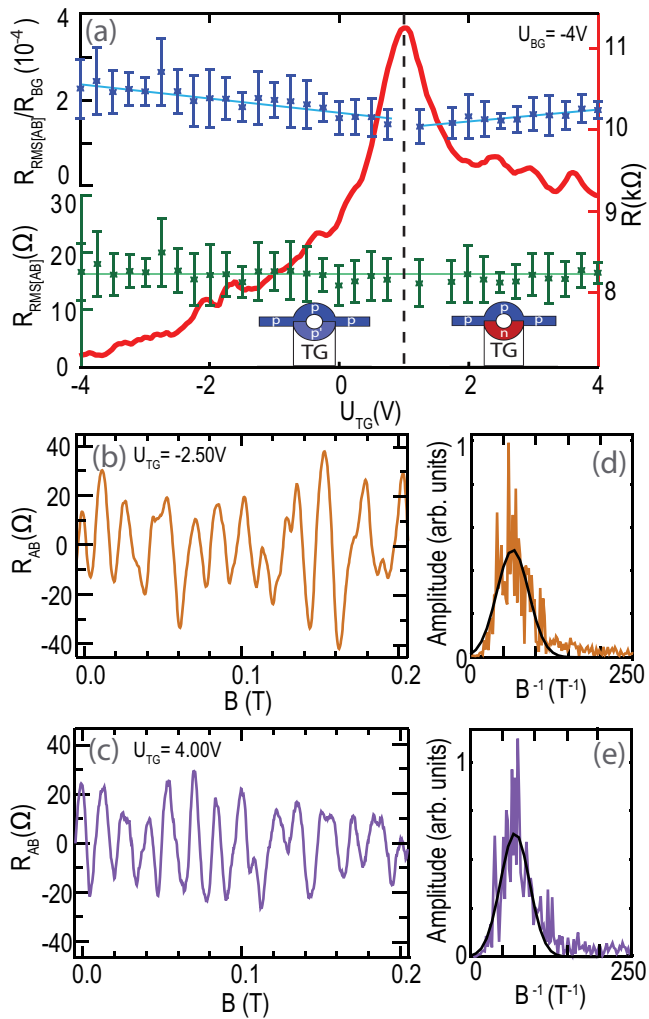} 
\par\end{centering}

\caption{\textit{Aharonov-Bohm oscillations dependent of the carrier type and
concentration. (a) The effective (green) and relative (blue) amplitude.
The asymmetry in the background resistance (red) is a direct proof
of the pnp junction in the ring. Insets are a schematic picture of
the charge carrier types in the ring influenced by topgate potential.
AB oscillations with the resistance background subtracted are shown
for unipolar hole- (b) and a bipolar electron-hole-measurement (c)
with a corresponding Fourier spectrum and fitting curves for both
oscillations (d-e). }}
\end{figure}
A typical measurement is presented in Fig.~3(a). It shows the resistance
of the ring versus the topgate voltage with a fixed backgate voltage
$U_{BG}=-4\,\mathrm{V}$ (red line), so that one side of the ring
and the leads have holes as charge carriers with a concentration of
$p=8.6\cdot10^{15}\,\mathrm{m}^{-2}$ due to the fixed backgate voltage, whereas the other side of the ring has a different charge concentration due to the influence of the topgate voltage.
The resistance of the topgate dependent CNP at $U_{TG}={1}\,\mathrm{V}$ with
$R=11.5\mathrm{\, k\Omega}$ is much lower in comparison to the backgate
dependent CNP (Fig.~1(a)) with approx. $R=40\,\mathrm{k\Omega}$.
This behavior is attributed to the small topgate-covered area which
is only one arm of the ring and is essentially smaller than the sum
of the other arm and the leads of the sample.

The CNP divides the graph into two regions with different charge carriers
in the second arm of the ring: holes on the left and electrons on
the right side. The asymmetry of the resistance is based on the non-perfect
Klein tunneling which depends strongly on the smoothness of the potential
step \cite{Falko2006,Katsnelson2006}. The higher resistance on the right side
of the graph shows the existence of the pnp junction created in one
arm of the ring. So Fig.~3(a) presents an unipolar system on the left
and a bipolar system on the right side of the CNP.

Whereas there is no clear observation of the AB oscillations at the
CNP, oscillations are observed away from the CNP.

Two sets of oscillations with a subtracted background are shown in Fig.~3(b) (unipolar) and 3(c) (dipolar). 
The charge carrier concentrations used in these two measurements are of similar magnitude but different polarity.
In both situations the absolute amplitude is quite similar.
The period of the oscillations is also not affected
by the unipolar or dipolar situation as seen from the Fourier spectra
shown in Fig.~3(d) and 3(e). The AB oscillations were analyzed for
a number of different topgate voltages. The RMS values of the absolute amplitude
is analyzed as described before and plotted versus the topgate voltage
in Fig.~3(a) as green dots. The measured RMS of the absolute amplitude is
more or less constant with an average value of $16.1\pm3.9\,\Omega$
as indicated in Fig.~3(a) by the horizontal green line. 
It does not change for different charge carrier type and concentrations.
The period of the oscillations is observed to be also constant in these topgate dependent measurements.

The relative amplitude is the absolute amplitude divided by 
the mean of the background resistance and can be used to characterize the visibility of the oscillations.
The relative amplitudes are shown as
blue dots versus the topgate voltage in Fig.~3(a) for our measurements. One observes a linear decrease
towards the topgate dependent CNP in both regions due to the overall
increase in resistance. The absolute value of the slope is higher
in the unipolar region than in the bipolar by a factor of $1.72$.
The minimum visibility in these measurements is approx. 0.2\% while
the maximum visibility reaches 0.3\%. This behavior is caused by the asymmetry
of the overall resistance while the actual absolute amplitude of the
oscillations being almost constant in both regions. In Ref. \cite{TheorieKleintunneling[13]}
the difference in visibility for the unipolar and bipolar situation
was explained by the tunneling of the charge carriers through the
junction and their interference with themselves. The resulting difference
in the relative amplitude is observed in our experiment, but the
astonishing fact remains that the absolute amplitude of the observed oscillations
is independent of whether holes interfere with holes or electrons
interfere with holes.

In conclusion we have reported AB oscillations in a monolayer graphene
ring with a period that fits the geometry of the ring. Our measurements
show that AB oscillations are possible in a ring system with a pnp
junction. No changes of period or amplitude are observed for this
dipolar regime.

We acknowledge discussions with P. Recher. This work was supported
by the DFG via SPP 1459, the excellence cluster QUEST and the NTH
School for Contacts in Nanosystems.
\newline


\end{document}